# A novel method of finding the dielectric constant of ceramic materials in microwave range with Dielectric Resonator Antenna setup using the HEM$_{11\delta}$ mode

Prithwiraj Ganguly[a], Vince Kumar[b], P. Maneesha[a], Saptarshi Ghosh[c], Somaditya Sen[a]*

[a]Department of Physics, Indian Institute of Technology Indore, 453552, India

[b]Department of Physics, Deenbandhu Chhotu Ram University of Science and Technology, Murthal, 131039, Sonepat, Haryana, India

[c]Department of Electrical Engineering, Indian Institute of Technology Indore, 453552, India

*Corresponding author; sens@iiti.ac.in

**Abstract**

A novel method to measure the dielectric constant ($\varepsilon_r$) of a material at GHz range has been introduced in this work. The dielectric measurement technique using the Micro-Strip Line coupled Dielectric Resonator Antenna (DRA) setup with simulation assisted results has been detailed. The resonance frequency corresponding to the HEM$_{11\delta}$ mode is extracted from the reflection coefficient ($S_{11}$ parameter) and utilized to measure $\varepsilon_r$ for DRA made from the material under investigation. Initially, ZrO$_2$ sample, with known $\varepsilon_r$, is used to verify the proposed method. An $\varepsilon_r$ value of ~17.5 has been experimentally obtained using the proposed method. Then, the measurement is done for a system with completely unknown $\varepsilon_r$ i.e., a composite of (1-x) BaTiO$_3$-(x)NiO, x=0, 0.05, 0.1, and 0.15. A detailed structural characterization study has been utilized for the BaTiO$_3$-NiO samples to understand the material properties at the fundamental level. The experimental value of the $\varepsilon_r$ in the GHz range was obtained as 22.9, 29, and 28.34 for x=0, 0.05, 0.1, and 0.15, using the proposed technique.

**Introduction:**

Dielectric materials play a pivotal role in modern technology, enabling the efficient and reliable operation of a wide range of devices and systems. Their unique insulating properties, ability to store and manage energy, and influence on signal transmission make them indispensable in electronics, telecommunications, energy storage and numerous other fields [1] [2]. For practical applications, it becomes essential to have information about the different electromagnetic properties of the material. Also, efficient techniques must be developed for easy and faster characterization.

The measuring techniques of a material's dielectric constant or permittivity ($\varepsilon_r$) in lower frequencies (up to MHz ranges) generally utilize the bridge circuit method [3]. While, at microwave frequencies, it can be broadly classified into: Waveguide methods [4], Resonance techniques [5], Perturbation methods [6], and other optical techniques [7] [4]. In the transmission line method, the complex $\varepsilon_r$ is calculated from the scattering parameters obtained by putting the sample in a coaxial or waveguide transmission line [4]. In the Resonance method, a resonant cavity with the sample inserted measures the shift in resonant frequency and change in quality factor, providing very accurate permittivity measurements for low-loss materials [8]. In these methods mostly the TE (Transverse Electric) or TM (Transverse Magnetic) modes are explored.

In the present work, the first of its kind of method is being proposed to estimate the $\varepsilon_r$ by calculating the resonance frequency corresponding to the HEM (Hybrid Electro-Magnetic) mode from the $S_{11}$ parameter for a Dielectric Resonator Antenna (DRA) made from the material concerned. DRA is a radiating non-metallic antenna, generally made of ceramics preferred due to less conductive loss, small size, availability of different coupling techniques etc. Depending on the shape and size, a DRA can radiate at a certain resonating frequency. A unique electromagnetic field is generated at these frequencies, which are not confined within the resonator itself. These are called non-confined modes [9]. These modes are utilized here to measure $\varepsilon_r$. Unlike the resonance or perturbation method, the introduced technique utilizes the resonance condition without using any metallic enclosure or cavity. To resonate in a particular mode, the feeding mechanism should be decided specifically. In this work, the DRA is fed with a Micro-Strip Line (MSL). This kind of coupling is efficient, simple, and easy to implement, where the DRA works like a finite waveguide placed on an open-circuited transmission line. This truncated section of dielectric waveguide acts as a resonator by creating a standing wave pattern on excitation with EM wave [10]. This new technique is advantageous due to the flexibility in choosing the size and the ease of measurement and analysis.

This work gives a detailed description of the introduced measurement technique and different precautions to be taken while designing the MSL-coupled DRA. Initially, a $ZrO_2$ sample is used to verify the proposed method. After that, the measurement is done for a completely unknown system, i.e., a composite of $BaTiO_3$-NiO. A detailed structural characterization study has been utilized for the $BaTiO_3$-NiO samples to understand the material properties at the fundamental level. After that, the introduced technique is utilized to get the $\varepsilon_r$ in the GHz range for these samples also.

**Experimental techniques:**

**Designing the MSL coupled DRA:**

A unique method is introduced to measure the $\varepsilon_r$ of the materials by using it as a DRA. This DRA is composed of a dielectric resonator (DR) placed on a MSL fed to a Sub-Miniature A-type (SMA) connector [Figure 2]. An EM-wave in the frequency range 1GHz-10GHz was fed to the SMA connector from a Vector Network Analyzer (VNA) (Anritsu S820E). Then by using the measured $S_{11}$ parameter (which shows a distinguishing dip at some frequency where most of the source energy gets transmitted and ultimately radiated via the DR) the resonance frequency is recognized and used for determining the $\varepsilon_r$.

The empirical formula for the resonance frequency of the hybrid $HEM_{11\delta}$ mode is given by [11]:

$$f = \frac{6.324c}{2\Pi r\sqrt{(\varepsilon r+2)}}[0.27 + 0.36\left(\frac{r}{2h}\right) + 0.002\left(\frac{r}{2h}\right)^2] \qquad (1)$$

This formula is known to be accurate in the range 0.4 < a/h < 6. Here, r, h, $\varepsilon_r$ and c represent radius, height, εr of the DRA and the velocity of the light in the free space respectively.

As the resonating frequency is dependent on the $\varepsilon_r$ of the material [equation 1], the selection of frequency range will lead to the fixing of $\varepsilon_r$ range for which the DRA will work. Here, the selected range is 1GHz to 10 GHz, so the $\varepsilon_r$ range will be ~15-80. The MSL design is made according to those specifications.

The antenna design was simulated and verified using ANSYS High Frequency Structure Simulator (HFSS) software. HFSS uses Finite Element Method (FEM) to solve electromagnetic equations for 3D structures working at high-frequency ranges (micrometer or millimeter wavelength). The signal was fed to the DR placed on a 50 Ω MSL fabricated on a Fiberglass-Reinforced epoxy-laminated sheet substrate (FR4, with $\varepsilon_{subs}$=4.4) with dimensions of 60 mm (length) × 60 mm (breadth) × 1.6mm (thickness) [Table 1]. The input impedance of 50Ω of the MSL was achieved with 35 mm (length) × 3 mm (width) MSL on the FR4 substrate. The MSL was experimentally fabricated with the help of a 3D printer (LPKF Protomat-S104) on one side of the FR4 substrate, while the opposite conducting side was retained as the ground plate with the same length × breadth as that of the substrate. Input impedance matching is crucial in determining the best energy coupling criteria [12]. Hence, the DR, MSL, and FR4 substrate dimensions were optimized using HFSS. One of the necessary conditions to achieve efficient coupling between the MSL and DR is $\varepsilon_{subs} \ll \varepsilon_{DRA}$ [13] is also taken care of while designing the DRA. The substrate, ground plane, and microstrip line parameters are tabulated in Table 1. In the simulation, a radiation box of air with a size of 120mm × 120mm × 50 mm is taken to calculate the far-field parameters. The dimension considers that all the box surfaces are at least $\lambda_o/4$ away from the radiating body ($\lambda_o$ is the wavelength corresponding to the resonance frequency) [14]. The values of radius and height of DR are initially optimized in HFSS to simulate the required design of the proposed DRA.

| Table 1: Optimized design parameters (in mm) | | | |
|---|---|---|---|
| DRA radius (r) | 5.5 | MSL width | 3 |
| DRA height (h) | 2.25 | Substrate thickness | 1.6 |
| MSL length | 35 | Substrate length × width | 60 × 60 |

The $S_{11}$ parameter of the fabricated MSL (without DR) was measured and simulated to verify the authenticity of the design [Figure 1]. The simulated and experimental values of the $S_{11}$ parameter are plotted in Figure 1(a) which shows similar trend. This can be considered as a perfect verification of the fabricated design. Kumar et.al. [13] mentioned, in the absence of DR, the EM waves create standing wave patterns by getting reflected from the open end. The HFSS simulation of the standing wave pattern corresponding to ~8.89 GHz is shown in Figure 1(b). After verifying the fabricated MSL, the position of the DRA on the MSL has been checked for maximum coupling.

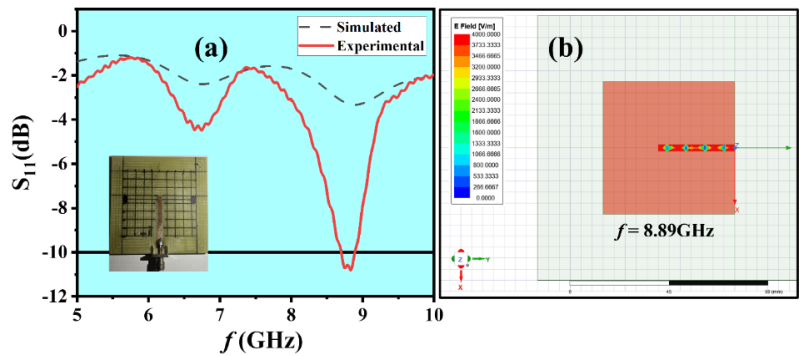

**Figure 1: (a) $S_{11}$ plot with the frequency of MSL without DRA (b) Simulated E-field magnitude showing standing wave pattern**

A Cylindrical Dielectric Resonator (CDR) offers three fundamental modes, $HEM_{11\delta}$, $TM_{01\delta}$, and $TE_{01\delta}$ [15]. Here, the first index denotes the number of full-period field variations in the azimuthal direction, the second one represents the no. of radial variations, and the third index represents the variation in the z-direction [11]. From the multipole analysis of the radiation pattern, it can be shown that the most prominent contribution comes from the lower order terms e.g. electric dipole or magnetic dipole (oriented in different directions) along with less weightage from the other higher order terms (i.e. quadrupole, octupole). The $HEM_{11\delta}$ mode is generated due to a magnetic dipole placed in the equatorial plane of the DR [15]. The feeding mechanism using this MSL and $\varepsilon_r$ of the DR ranging between ~15-80, yields a prominent $HEM_{11\delta}$ mode in the frequency range 1GHz to 10GHz. This resonating frequency is then utilized to determine the material's exact $\varepsilon_r$ in the GHz range.

A CDR, when positioned near the MSL, effectively couples the magnetic field lines associated with it. The key to successful coupling lies in the critical positioning of the CDR on the MSL. The length of the open end of the MSL under the DRA, known as the overlapping distance ($l_o$), plays a crucial role in matching impedance and enabling coupling between the DR and the MSL for a specific mode [15]. By properly adjusting the overlapping distance ($l_o$), the MSL feeding technique offers the advantage of significantly improving coupling. In this context, optimal coupling is achieved when $l_o$ equals 2.5 mm. It has been prominently observed that symmetric placement of the CDR on the axial line of the MSL is key to minimizing modal impurity and diminishing the presence of higher-order multi-polar components [16], [17]. As a result, great care has been taken throughout the experiment to ensure the center of the DR is precisely positioned on the axial line of the MSL.

However, along with the $HEM_{11\delta}$ mode some other higher-order modes may also resonate in the given frequency interval. In other words, properly detecting the $HEM_{11\delta}$ mode is essential in this technique. The $HEM_{11\delta}$ mode is verified by simulating the E-field and H-field distribution within the DRA. The broadside radiation gain pattern is also proof of the same mode and hence, has been studied [16]. Confirmation of the mode using field distribution, and the radiation pattern enables the correct detection of the resonant frequency of the $HEM_{11\delta}$ mode.

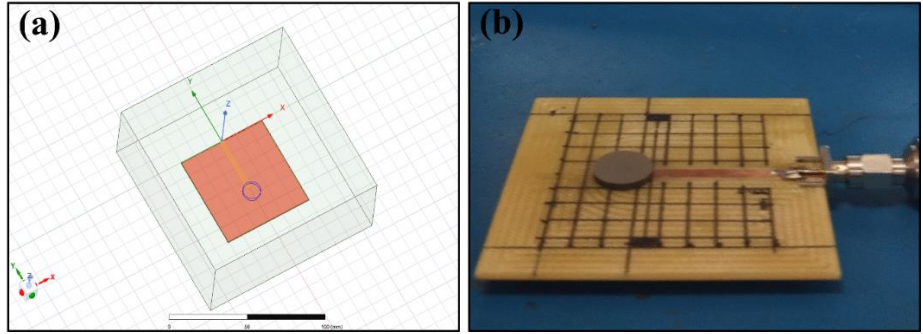

**Figure 2:** (a) Simulated and (b) Fabricated DRA with Micro-Strip Line (MSL)

**ZrO$_2$ DR preparation**

ZrO$_2$ (99% pure) is used to make DRA for the proposed methodology verification. X-ray diffraction [Supplementary Figure 1] reveals the P2$_1$/c phase (Monoclinic phase) of the ZrO$_2$ and the structure is confirmed with the Raman spectroscopy [Supplementary Figure 2]. The ZrO$_2$ powders with Polyvinyl Alcohol (PVA) are pressed into cylindrical pellets of diameter ~13 mm, using a uniaxial hydraulic press (with 2 tons of pressure for 2 mins). PVA is adding to the powders as binder which helps to makes compact pellets. The pellets are pre-heated at 600 $^0$C for 6 hours to remove the PVA. Then the pellets are sintered at 1250 $^0$C for 5 hours with a heating and cooling rate of 5 $^o$C/min. After sintering the dimension of the ZrO$_2$ pellets obtained with radius r= 5.91mm and height h= 2.362 mm.

**BTO-NiO sample preparation and characterization tools:**

BTO-NiO composite powders were prepared using commercially available nano-sized particles of BaTiO$_3$ and NiO. The molecular weight percentage of NiO in (1-x) BaTiO$_3$-(x)NiO varied as x=0, 0.05, 0.1, and 0.15. These samples are named as S0 (x=0), S1 (x=0.05), S2 (x=0.1), and S3 (x=0.15). Stoichiometric combinations of pure BTO and NiO powders were mixed and ground for three hours in an agar mortar. The process ensures the homogeneity of the mixture. The powders of each composition were pressed into three cylindrical pellets of diameter 10 mm, using a uniaxial hydraulic press (with 2 tons of pressure for 2 mins). Two pellets were prepared with PVA to ensure proper particle adhesion. The third pellet was prepared without adding PVA to avoid unwanted components from remnants of the PVA. All the pellets were sintered at 1250$^0$C for 5 hours with a heating and cooling rate of 5 $^o$C/min. The sintered pellets for each sample that were pressed without PVA were ground into powder for basic characterizations (XRD and

Raman). The PVA-added pellets were used for morphology tests using FESEM. The same pellets were used for dielectric property measurements.

A Bruker D2-Phaser X-ray diffractometer has been used to get the X-ray diffraction (XRD) data for structural analysis. A Cu $K_\alpha$ source ($\lambda$ =1.54 Å) at 30 kV and 10 mA at the sweep rate of 0.5/min is used to do XRD in the range of 20° to 80°. The phonon modes were studied from room temperature Raman spectroscopy using a Horiba-made Lab-RAM HR Raman spectrometer (spectral resolution of 0.9 cm$^{-1}$). The illumination source was a He-Ne LASER of wavelength 632.8 nm. A CCD detector is used in backscattered mode along with 600 grating. A Supra55 Zeiss Field Emission Scanning Electron Microscope (FESEM) was used to study the sintered pellets morphology and to obtain Energy dispersive X-ray (EDX) data. A metallic gold deposition (~ 5 nm) by sputtering techniques was required for charge drainage from the surface.

The same sintered pellets were used to estimate the dielectric properties. Electrodes were painted using silver paste on both sides of the pellets for the dielectric measurements. Curing of the silver electrodes was performed at 550°C for 30 minutes to ensure proper adhesion of the Ag to the pellet surface. A broadband (100 Hz to 1 MHz) Newton's 4th Ltd. dielectric spectrometer was used for the dielectric measurement. The spectrometer is equipped with a phase-sensitive multimeter having a signal strength of 1 V rms. All measurements are conducted at room temperature.

To analyze the resonating properties and the dielectric characteristics (at GHz range), cylindrical pellets of the materials were prepared with almost similar dimensions for S1, S2, and S3. These CDRs were fabricated using a 13 mm dies pellet maker. Sintering of the CDRs was performed at 1250°C using a similar protocol as the pellets for dielectric measurement. The intention was to keep the dimensions constant for all three materials (S1, S2, and S3). Still, due to differences in their responses to sintering, the size of the pellets has varied nominally from sample to sample. The radius, $r_{DRA}$ varied from 5.49mm (S1) to 5.49mm (S2) and 5.72mm (S3), while the height $h_{DRA}$ varied from 2.62mm (S1) to 2.26mm (S2) and 2.25mm (S3).

**Results and Discussion:**

**Determining the $\varepsilon_r$ of ZrO$_2$ in the 1GHz to 10GHz range:**

The value of $\epsilon_r$ of ZrO$_2$, which was previously determined and discussed in the literature as ~17-25 [17] [18] depending on the phase and the particle size. A DRA of ZrO$_2$ of dimensions r

= 5.91 mm and h = 2.362mm is prepared. The ZrO$_2$ DR was placed on fabricated MSL with l$_0$=2.5mm. The S$_{11}$ spectra revealed a resonance frequency at $f$ = 8.38 GHz [Figure 3(a)]. The obtained impedance value is 54.18+j 2.88 Ω at the resonance frequency [Figure 3(b)]. This reflects an excellent energy coupling between the DR and the MSL.

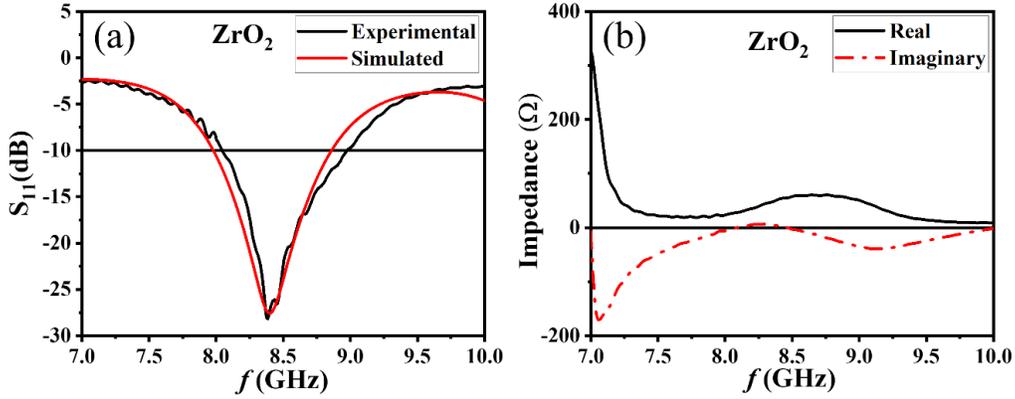

**Figure 3: (a) S$_{11}$ plot with frequency for ZrO$_2$ (b) Real and Imaginary impedance plot with frequency**

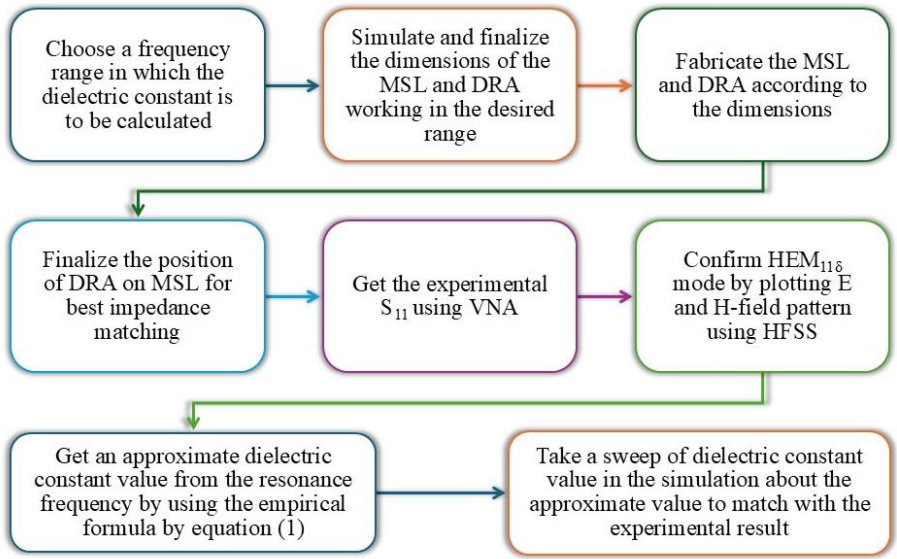

**Figure 4: Flow-chart of the steps to measure the ε$_r$ using the introduced technique**

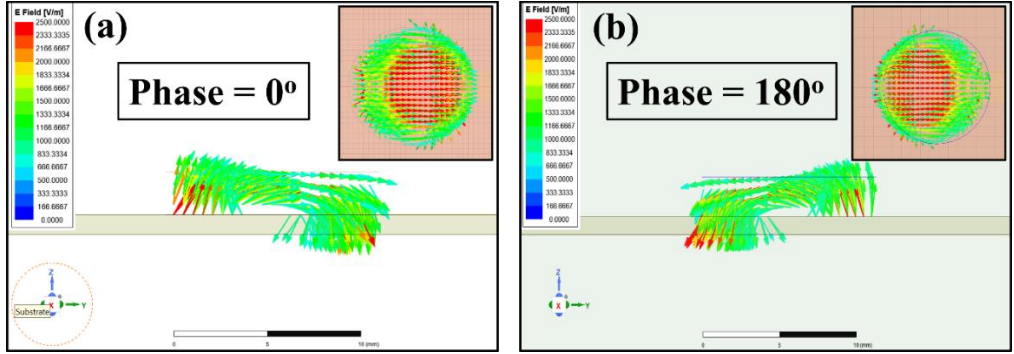

**Figure 5: E-Field at (a) Phase=0° and (b) Phase=180° within the DRA at $HEM_{11\delta}$ mode (top view inlet)**

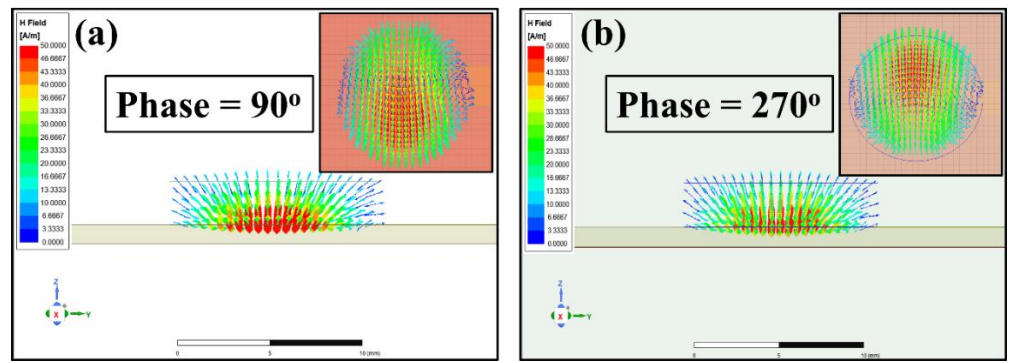

**Figure 6: H-field at (a) Phase=90° and (b) Phase=270° within the DRA at $HEM_{11\delta}$ mode (top view inlet)**

The electric and magnetic field patterns are simulated to analyze the mode generated at the resonance frequency. [Figure 5] and [Figure 6] show the different field distributions at the resonant frequency (8.38 GHz). This distribution confirms that it was $HEM_{11\delta}$. This is the hybrid mode with the lowest frequency [10]. In the present case, the dipole was placed in the X-Y plane, oscillating along the X-axis. The magnetic field was strongest in the equatorial plane, whereas the electric field was strongest at the top surface. The electric field lines curled around the magnetic field lines. The side view of the E-field at phase = 0° [Figure 5(a)] and phase = 180° [Figure 5(b)] revealed that the electric field vectors were getting directed in the opposite direction. From the top view, it is observable how the maximum magnitude of the E and H fields are oscillating (denoted with red-colored vectors) about the center. A similar analysis can be inferred for H-field. It is to be noted that the E-field and the H-field maximas are shifted by a 90° phase difference in the time domain. The H field pattern at phase=90° [Figure 6(a)] and at phase=270° [Figure 6(b)] are plotted.

To further justify the generated HEM$_{11\delta}$ mode at the resonant frequency, the gain and the directivity are analyzed at the resonating frequency. The gain and directivity plots are shown in [Figure 7]. The maximum gain obtained is 5.8 dB. It is observed that the proposed prototype has an isotropic gain in the broadside direction corresponding to the HEM$_{11\delta}$ mode [11].

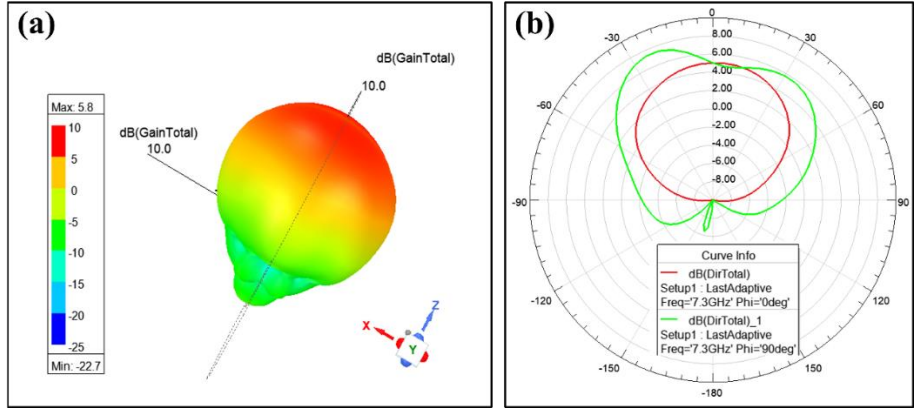

**Figure 7: (a) Gain and (b) Directivity plot**

To match the simulated S$_{11}$ with the experimental value, the approximate value $\varepsilon_{approx}$ ~ 18.9 was calculated from the empirical formula. It is used as an initial value of $\varepsilon_r$ for the simulation to start with. The experimental dimension of the fabricated DRA is used to obtain a simulated S$_{11}$ parameter. The simulation was targeted to match the resonant frequency with the experimental resonant frequency by varying the $\varepsilon_r$.

Thereafter, the value is refined by sweeping $\varepsilon_r$ in a range around the approximate value $\varepsilon_{approx}$ to match the experimental resonance frequency. The obtained refined value using HFSS is $\varepsilon_r$ ~17.5 [Figure 3a]. This value is an excellent match with an experimental work reported elsewhere (~17.9) [18].

The ZrO$_2$ could have been enough to prove the technique itself. However, to generalize and demonstrate the versatility of the method for all types of samples, including solid solutions and composites of ceramics, an attempt is made hereafter by studying a more complex system of a combination of a major tetragonal BTO phase and a cubic NiO phase. For lower concentrations of NiO, a transformed BTO phase is explored, while for higher concentrations, a composite phase is analyzed. The experimental observations for these samples and some basic structural analyses are presented in the next section.

**Determining the ε_r of BTO-NiO composite samples:**

**XRD analysis:**

A mixture of (1-x) BTO-(x)NiO has been chosen to prepare a sample. From the XRD studies, the sintered samples revealed a composite form of BTO and NiO for S2 and S3, while S0 and S1 revealed a pure BTO structure. A major BaTiO$_3$ tetragonal phase with a *P4mm* space group was observed for all samples with [Figure 8] some amount of NiO cubic [*Fm-3m*] phase in S2 and S3. The samples exhibited well-defined sharp peaks, confirming the crystalline nature of the samples. The intensity of the NiO peaks is found to increase from S2 to S3, confirming the composite nature. Some nominal non-stoichiometric impurity peaks are observed in the 2θ ~ 27°-31° range in all the samples corresponding to Ba$_2$TiO$_4$, Ba$_4$Ti$_{12}$O$_7$, Ba$_2$Ti$_2$O$_5$, Ba$_6$Ti$_{17}$O$_{40}$ phases [19].

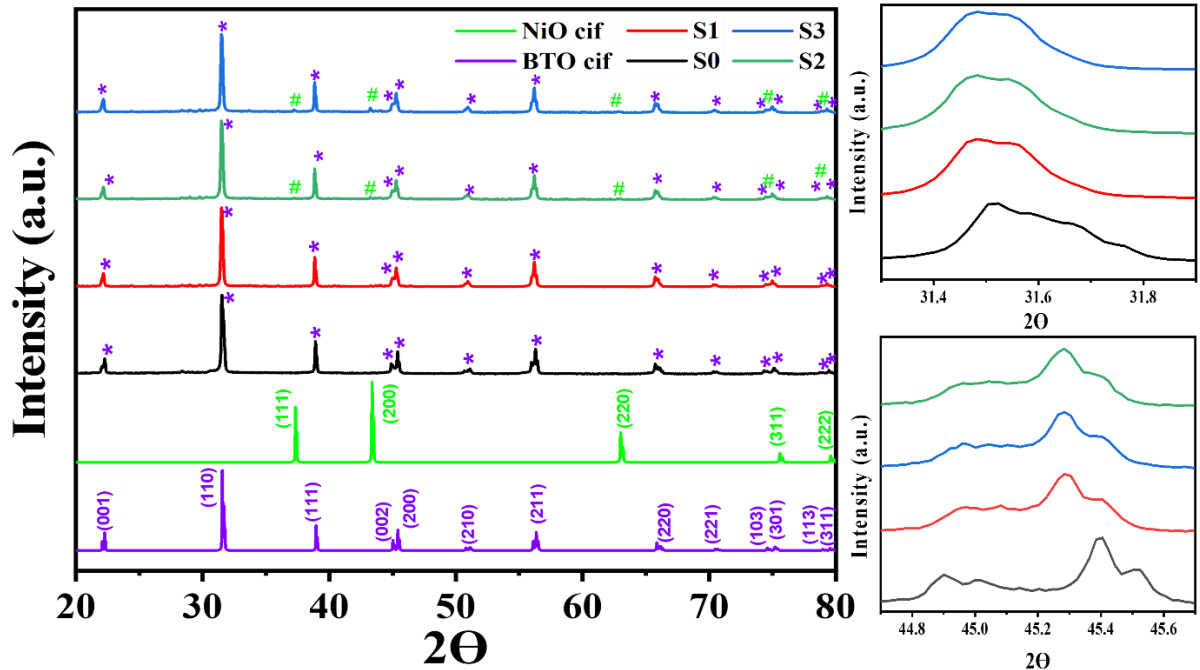

**Figure 8: (a) X-ray diffraction plot of synthesized samples [BTO peaks (\*) and NiO peaks (#)]. (b) Zoomed image of the highest intensity (100) peak. (c) Zoomed image of the (200) peak.**

In the ABO$_3$ perovskite structure, the BO$_6$ octahedra plays the primary role in ferroelectricity or polarization. In the tetragonal phase of BaTiO$_3$, the central Ti ion does not remain at the center of the TiO$_6$ octahedra, leading to an offset of positive and negative charge centers. This, in return, gives rise to a macroscopic spontaneous polarization P$_s$ along the [001], resulting in ferroelectricity. Centrosymmetric position of Ti leads to the cubic crystal structure and reduction of $P_s$. The splitting of two peaks (002) at ~44.9° and (200) at ~ 45.4° confirmed the non-centrosymmetric tetragonal phase of $BaTiO_3$ [Figure 8(c)]. The highest intensity was of the (110) plane at ~ 31.52°, representing a preferred crystallinity in this orientation for all the fabricated polycrystalline BTO. Slight asymmetry was observed for the (110) plane, indicating the tetragonality of these samples [Figure 8(b)] [20]. Two possibilities emerged due to the addition of NiO to $BaTiO_3$: (a) Ni doping in $BaTiO_3$ lattice and (b) a $BaTiO_3$-NiO composite formation. The sintering temperature of $BaTiO_3$ is reported in the literature to be ~ 1350 °C [21]. Addition of NiO has reduced the sintering temperature. The double peaks (002/200) representing the tetragonal nature seemed to converge with NiO incorporation, thereby losing the tetragonality. Such a nature is generally the signature of doping. Hence, one shouldn't neglect such a possibility in these samples. Therefore, the phase percentage of both NiO and $BaTiO_3$ were estimated from Rietveld refinement. NiO peaks are not present in S1, indicating the entire NiO consumed by the BTO lattice, representing a complete doping. However, traces of NiO are observed in S2 (~1.57%) and S3 (4.63%). This indicates that NiO is highly soluble in the BTO lattice with a solubility limit of ~10%. Hence, the composite system exists only in the S2 and S3 samples.

In these sintered mixtures of BTO and NiO possibly substitute either a B-site or an A-site or both A and B site of BTO in these samples. Substitution of any one ion in BTO will lead to the imbalance of Ba:Ti ions ratio 1:1 creating cationic defects. Such cationic defects are acceptable in lower doping percentages but are structurally less logical for higher (>2%) doping. In search of different possibilities, it was observed that there are reports for Ni-doped BTO (> 2%) with Ni replacing Ti at the B-site. Such replacements generally have been reported to trigger a hexagonal *$P6_3/mmc$* phase of BTO for Ni presence of ~2% [22]. On the other hand, there are other reports of $Ni^{3+}/Ni^{4+}/Ni^{2+}$ ions replacing $Ba^{2+}$ at the A-site [23]. Such a substitution instigates a non-centrosymmetric tetragonal to a more centrosymmetric cubic BTO (Fm-3m) structure [23]. Note that both Ni and Ti have 3d electrons in the outermost shells. Hence, it is more probable that Ni should be substituted for Ti. However, this should generate hexagonal phases. This was not observed in these samples. On the other hand, a loss of tetragonal phase in these samples is observed, which leads one to believe that Ni is substituting Ba in the modified BTO lattice. The fact that neither the hexagonal phase nor the cubic phase has been revealed hints at a competition of these two mechanisms which allows the phase to remain tetragonal. Hence, one indication of such behavior is correlated to a simultaneous substitution of both A-site and B-site by Ni. The following analysis is designed to prove this hypothesis.

The crystal radius of $Ba^{2+}$ [XII] is 1.75 Å, while that of $Ti^{4+}$ [VI] is 0.745 Å. Crystal radius of Ni depends on the valence state and the coordination of the ion. Ni should be XII-coordinated for an A-site substitution, while a B-site substitution should be VI-coordinated. From Shannon radii information $Ni^{2+}$ [VI] is ~0.83 Å, while $Ni^{3+}$ [VI] is ~0.7 Å for low spin state and 0.74 Å for high spin state. The lesser probable $Ni^{4+}$ [VI] is ~ 0.62 Å and can have only a low spin state [24]. From these values, one can observe that $Ni^{2+}$ is larger than the $Ti^{4+}$. However, $Ni^{3+}$ seems more comparable with $Ti^{4+}$, whereas $Ni^{4+}$ is much smaller. The valence state of Ni in NiO is $Ni^{2+}$. Hence, incorporating such an ion will likely generate expansive pressure in the lattice. This expansive pressure might be the probable reason for a transformation to the hexagonal phase. However, $Ni^{2+}$ in an A-site should be XII-coordinated for which information is not available in the Shannon radii table. An approximation using the extrapolation of the values of IV, V, and VI-coordinated radii enables one to obtain a crystal radius of $Ni^{2+}$ [XII] comparable to $Ba^{2+}$ ion. This type of substitution should in principle lead to the changes of lattice constants. Rietveld refinement [Supplementary Figure 3(a)] reveals modifications of the lattice constant with NiO incorporation in BTO as tabulated in Table 2. The c/a ratio which is a signature of the tetragonality is found to be reduced in all samples in comparison to BTO.

**Raman spectrum analysis:**

| Table 2: structural parameters obtained after Rietveld refinement ||||||
| (1-x) BTO-(x)NiO | a=b (Å) | c (Å) | BTO cell volume | Tetragonality (c/a) | Phase percentage ||
| | | | | | $BaTiO_3$ | NiO |
| --- | --- | --- | --- | --- | --- | --- |
| S0 | 3.9957 | 4.0347 | 64.4152 | 1.0098 | 100 | - |
| S1 | 3.9989 | 4.0245 | 64.3560 | 1.0064 | 100 | - |
| S2 | 3.9981 | 4.0252 | 64.3414 | 1.0068 | 98.43 | 1.57 |
| S3 | 3.9995 | 4.0245 | 64.3774 | 1.0063 | 95.37 | 4.63 |

Raman spectroscopy is known to be a very sensitive tool for obtaining information about changes in the local symmetry of the structure. Any small changes in structure due to doping should be reflected in the Raman spectra better than any other technique. The Raman spectra of pure BTO powders reveal prominent peaks corresponding to the E(TO) phonon mode at ~40 $cm^{-1}$

, an interference feature at ~80 cm$^{-1}$, a broad A$_1$(TO) mode at 275 cm$^{-1}$, a combined sharp B$_1$ mode and E(TO+LO) mode at 307 cm$^{-1}$, a combined A$_1$(TO) mode and E(TO) mode at ~520 cm$^{-1}$ and finally a combined A$_1$(LO) and E(LO) mode at 717 cm$^{-1}$ [25], [26]. The nearby A$_1$(TO) and the E(TO) modes ~ 520 cm$^{-1}$ appear to be convoluted and appear as a broad asymmetric peak. However, with doping this convolution reduces and these two can be visibly deconvoluted [27].

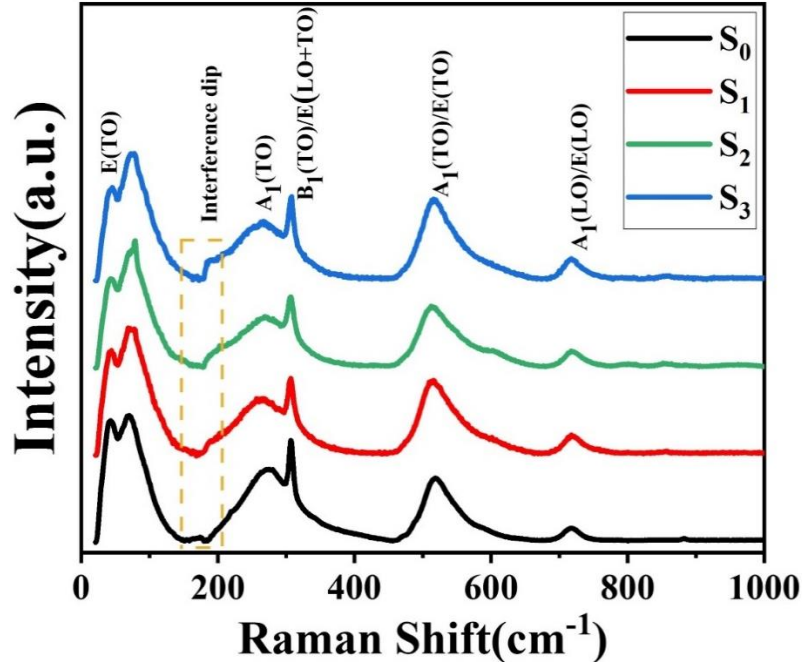

**Figure 9: Raman spectrum of all samples**

The peaks at 307 cm$^{-1}$, 520 cm$^{-1}$, and 717 cm$^{-1}$ arise due to the non-centrosymmetric nature of the TiO$_6$ octahedra and a signature of the tetragonality of the phase. Hence, a major tetragonal phase in all these samples can be confirmed and is at par with the XRD data. There are no Raman peaks corresponding to the hexagonal phase seen in any of the samples. The intensity of the strongest peak at 307 cm$^{-1}$ is not altered significantly, indicating the presence of tetragonality in all the samples. The prominence of the A$_1$(TO) mode at ~540 cm$^{-1}$ in the S1, S2, and S3 samples compared to S0 is correlated to the relative vibrations of the O-cage with the central Ti atom, thereby varying the polarity of the octahedron in the doped samples. This is valid proof of the presence of Ni at the Ti-site. The fact that Ni is also substituting Ba at the A-site can be revealed from the enhancement in the interference dip at 180 cm$^{-1}$ [interference due to

A(TO)+A(LO)+E(LO)+E(TO) (corresponding to the out of phase vibration of Ba atoms and TiO$_6$ octahedra) [25] and the broadness of the A$_1$(TO) mode at 275 cm$^{-1}$. Ni at the A site has a weaker bonding with the basal O, thereby reducing the strength of the A-O bond, which is responsible for the distortion of the lattice in terms of the displacement of the O-cage away from the centrosymmetric position. The reduction of the intensity of these vibrations is a signature of the loss of the much-required distortions essential for ferroelectricity. Hence, these results are essential proof that the ferroelectricity in the doped materials has been compromised to some extent without absolutely losing it. The shifts in the Raman peaks observed for the prominent tetragonal peaks are tabulated in Table 3. One notable observation is the broadening of the most prominent E(TO) mode at 40 cm$^{-1}$, A$_1$(TO) mode at 275 cm$^{-1}$, and B$_1$(TO+LO) mode at 307 cm$^{-1}$ along with the asymmetric broadband ~520 cm$^{-1}$. This is a clear indication of the increasing positional disorder arising due to the presence of Ni in the BTO lattice [28].

| Table 3: Raman peak position (cm$^{-1}$) | | | | | |
|---|---|---|---|---|---|
| (1-x)BTO-(x)NiO | E(TO) | A1(TO) | B1(TO)/E (LO/TO) | A1(TO)/E(TO) | A1(LO)/E(LO) |
| S0 | 42.82 | 275.52 | 307.99 | 518.04 | 717.42 |
| S1 | 44.11 | 268.01 | 307.99 | 515.62 | 717.42 |
| S2 | 42.82 | 268.01 | 307.99 | 515.62 | 717.42 |
| S3 | 45.41 | 266.75 | 307.99 | 515.62 | 717.42 |

**SEM and EDX analysis:**

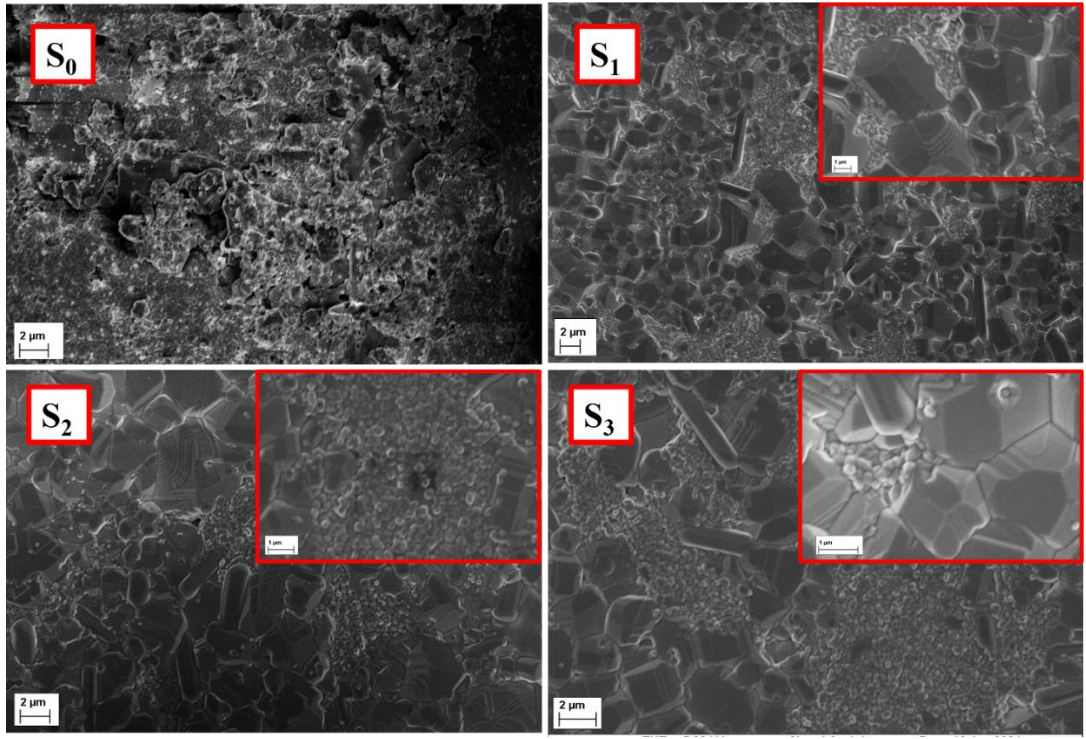

**Figure 10: Field Emission Scanning Electron Microscopy (FESEM) Images of sintered pellets**

SEM images of S1, S2, and S3 samples [Figure 10] reveal dense pellets with a minimal porous nature. These samples are, therefore, perfect for DRA applications. The morphology of the samples revealed a combination of two different formations, cuboid and granular morphology. However, for S0, proper sintering was not possible at 1250 ºC, thereby revealing a porous surface. Incorporation of NiO in BTO increases the granular morphology due to the formation of micro regions. Hence, NiO plays an important role in reducing the sintering temperature.

As the difference in percentage between the granular phase [Figure 11a] and the cuboid phase [Figure 11b] was maximum, EDX mapping was done for the S3 sample [EDX spectrum is shown in supplementary, Figure 4]. The sample revealed inhomogeneity of the constituent elements revealing specks of concentrated Ni-rich dots in the 2D maps of the samples. The homogeneity was obtained for the Ba and Ti elements.

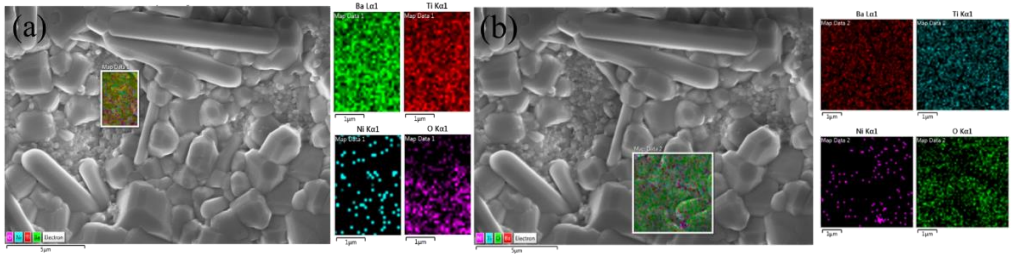

**Figure 11: Elemental Mapping on (a) Granular morphology (b) Cuboid morphology**

**Dielectric study:**

The complex permittivity, $\varepsilon_r^*(\omega) = \varepsilon'_r(\omega) - i\varepsilon''_r(\omega)$, where the real $\varepsilon'_r(\omega)$ part represents the energy storing capability of the material and the imaginary $\varepsilon''_r(\omega)$ part represents the energy dissipation in the form of heat. Dielectric spectroscopic studies have been done in the frequency range from 1Hz to 1 MHz. In this range $\varepsilon'(\omega)$ is calculated with the following equation, $\varepsilon'_r(\omega) = C_p d / \varepsilon_0 A$, where $C_p$, d, A, and $\varepsilon_0$ are the measured capacitance, thickness, area of the sample and $\varepsilon_r$ of the free space respectively. The room temperature frequency-dependent permittivity and loss tangent is plotted in Figure 12.

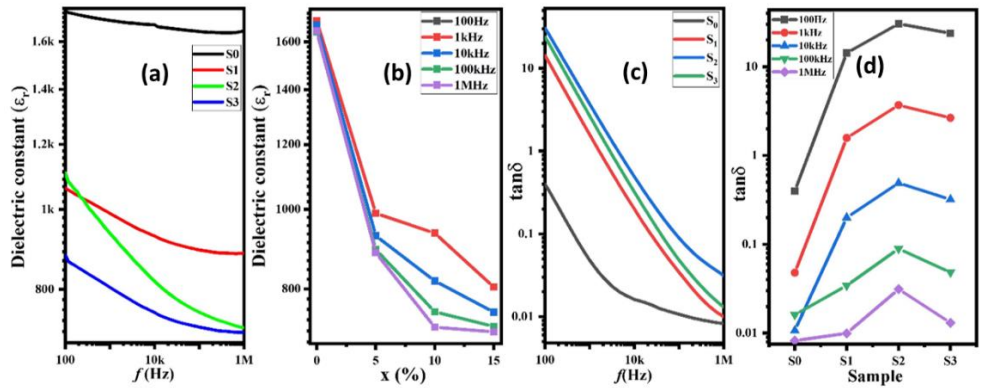

**Figure 12: Room temperature (a) Dielectric constant vs freq. (b) Dielectric constant vs sample composition (c) tan delta vs freq. (d) tan delta plot for different samples**

The permittivity reduces drastically in the entire frequency range (100 Hz to 1 MHz) from S0 (1800 at 1kHz and 1700 at 1MHz) to the modified samples S1 (1000 at 1kHz and 900 at 1MHz), S2 (950 at 1kHz and 700 at 1MHz), and S3 (800 at 1kHz and 700 at 1MHz) [Figure 12(a)]. The $\varepsilon_r$ of the samples depends mainly on four different contributions i.e., Ionic polarization includes space charge and interface polarization ($\varepsilon_i$), dipolar polarization ($\varepsilon_d$), atomic polarization ($\varepsilon_a$), and electronic polarization ($\varepsilon_e$). The response from atomic and electronic polarization becomes active at a frequency range of ~ $10^{16}$ and $10^{13}$ Hz, respectively. According to the Maxwell-Wagner interfacial polarization model, the contribution from space charge polarization remains effective ~ <10kHz. At the higher frequencies, the dipoles cannot follow alteration of the applied ac electric field; thus, though at lower frequencies $\varepsilon_i$, $\varepsilon_d$, $\varepsilon_a$ and can contribute to $\varepsilon'(\omega)$ and at higher frequencies, only $\varepsilon_e$ plays the role. This leads to a significant drop in the $\varepsilon_r$ at the high-frequency range.

All of the samples follow the Universal Dielectric Response (UDR) model [Supplementary Figure 5]. The reported value of $\varepsilon_r$ of pure BTO is ~100 **[29]** and that of NiO is ~3-5 [30] in the microwave region. Hence, the composite of BTO and NiO can expect an $\varepsilon_r$ value in the range between these values.

The tan$\delta$ increases considerably in the frequency range (1kHz to 100 kHz) for S0 (0.0479 at 1kHz and 0.01607 at 100 kHz) to the modified samples S1 (1.575 at 1kHz and 0.0343 at 100 kHz), and S2 (3.697 at 1kHz and 0.0891 at 100 kHz), and thereafter reduces for S3 (2.653 at 1kHz and 0.0483 at 100 kHz) [Figure 12]. Hence, for all the samples a reduction of tan$\delta$ is observed with increasing frequency. Therefore, for frequency in the range of GHz the tan$\delta$ can be assumed to be low, which is a required condition.

With the incorporation of Ni in the BTO lattice there is a change of the lattice composition, which seems to reduce $\varepsilon_r$. Also, the incorporation of the NiO forms the micro-regions in between the modified BTO grains. This will lead to the proper sintering of the prepared samples even at low temperature. Thus, the properties analyzed for the BTO-NiO solid solution (S1) and composites (S2 and S3) reveal that NiO incorporation reduces the sintering temperature with a prominent reduction in the permittivity and low loss tangent.

The experimental $S_{11}$ parameter for S1, S2, and S3 revealed resonance for the HEM$_{11\delta}$ mode at 7.381GHz, 7.2865GHz and 7.2865GHz respectively. The impedance at the resonance frequency for S1, S2, and S3 is 53.47-j4.08 $\Omega$, 56.93-j9.04 $\Omega$, and 52.71-j4.08 $\Omega$ respectively [Figure 13(d)]. This reflects that the energy coupling between the MSL and the DRA is up to the mark.

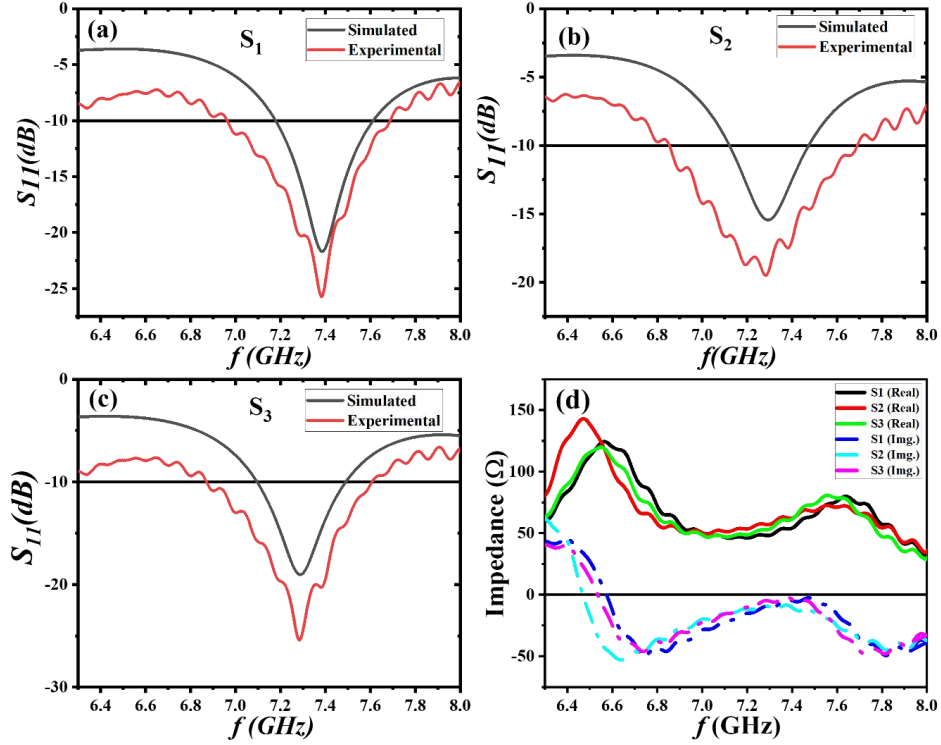

**Figure 13:** Simulated and experimental $S_{11}$ plot of (a) S1 (b) S2 (c) S3 sample. (d) Frequency-dependent Impedance plot for S1, S2, and S3

The approximate $\varepsilon_{approx}$ calculated from the experimental resonant frequency values for S1, S2, and S3 were 22.9, 29, and 28.34, respectively. Further, a simulation was performed to match the resonance frequency, and the refined values of $\varepsilon_r$ were found to be 24.2, 32.3, and 29.5 for S1, S2 and S3 respectively.

**Conclusion**

An indirect method for the measurement of $\varepsilon_r$ at GHz range has been proposed using the Dielectric Resonator Antenna setup. The material under investigation has been made as a DRA and from the reflection coefficient ($S_{11}$) measurement the dielectric constant of the material has been calculated. $HEM_{11\delta}$ mode of DRA is used to determine the dielectric constant. The occurrence of $HEM_{11\delta}$ was confirmed using the E and H fields and radiation pattern analysis using HFSS. The proposed method has been verified by using a known $\varepsilon_r$ material $ZrO_2$. The value of $\varepsilon_r$ of $ZrO_2$ has been obtained as ~17.5 using the proposed method in agreement with the earlier reports. A composite system of (1-x) $BaTiO_3$- (x)NiO (x=0, 0.05, 0.1, and 0.15) with unknown $\varepsilon_r$ is used for the further verification of the introduced technique. The value of $\varepsilon_r$ of these samples calculated from the obtained experimental resonant frequency values are 24.2, 32.3, and 29.5 for x= 0.05, 0.1, 0.15 respectively.

**Acknowledgment**

The authors would like to acknowledge the Sophisticated Instrumentation Facility, IIT Indore For the FE-SEM and Raman Spectroscopy measurement. The Authors would like to acknowledge the Department of Science and Technology (DST), Govt of India for providing the funds (DST/TDT/AMT/2017/200). The authors also acknowledge the Department of Science and Technology (DST), Govt. of India, New Delhi, India, for providing FIST instrumentation fund to the discipline of Physics, IIT Indore, to purchase a Raman Spectrometer (Grant Number SR/FST/PSI-225/2016). MP would like to thank the Ministry of Education, Government of India, for the Prime Minister Research Fellowship (PMRF).

# Supplementary Datasets
## A novel method of finding the dielectric constant of ceramic materials in microwave range with Dielectric Resonator Antenna setup using the HEM$_{11\delta}$ mode

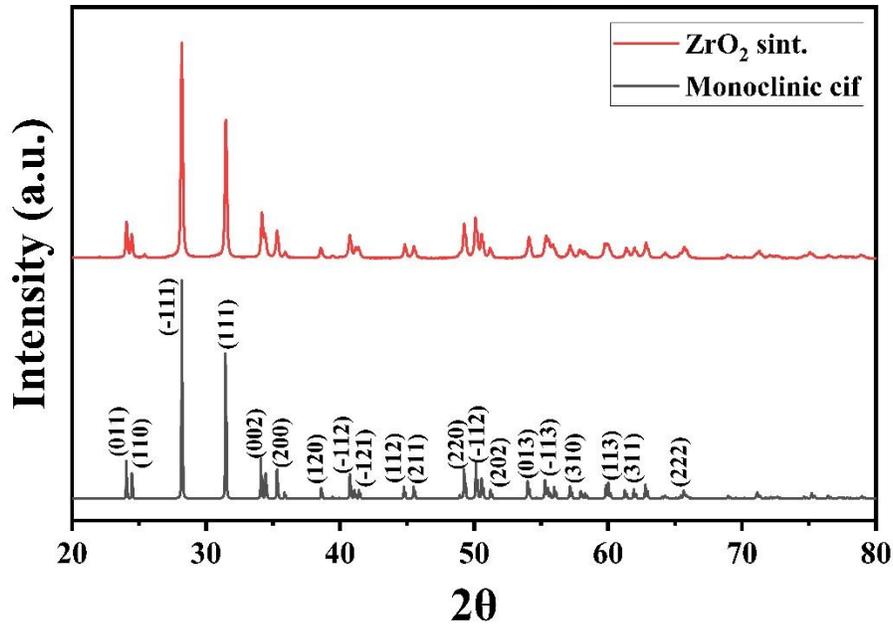

**Figure 1: X-Ray Diffraction pattern of ZrO$_2$**

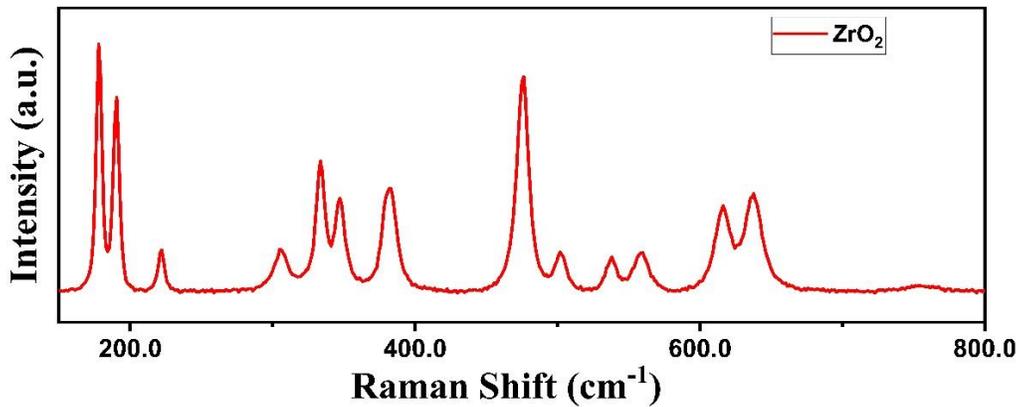

**Figure 2: RAMAN spectra of ZrO$_2$**

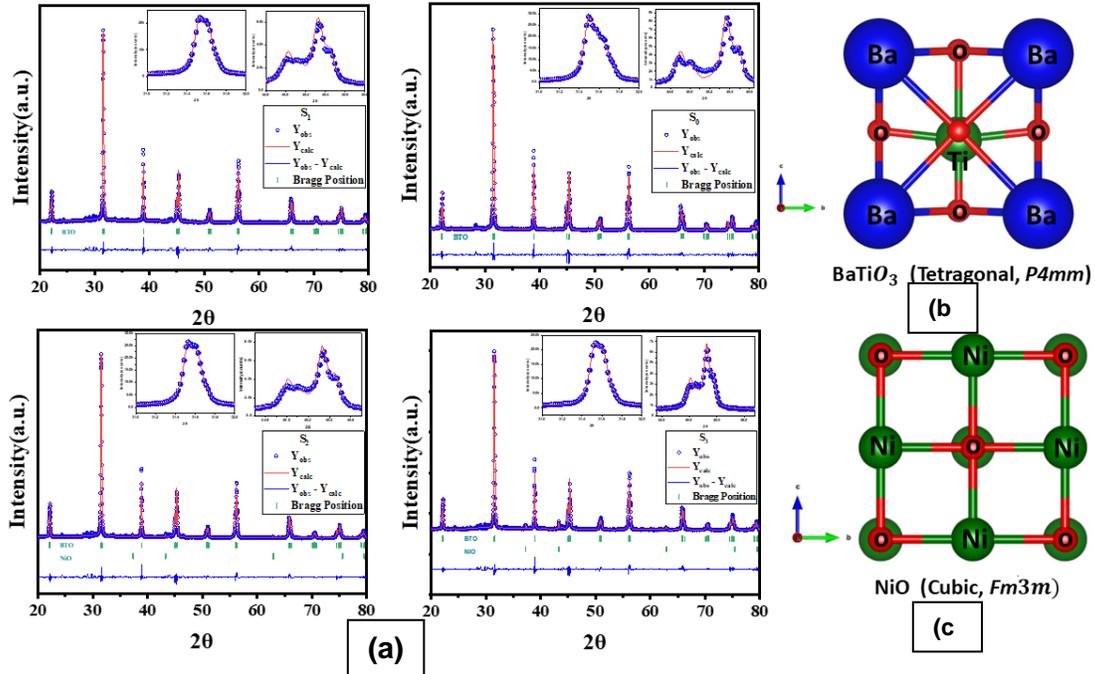

**Figure 3: (a) Rietveld measurement plots of the samples (b) BaTiO$_3$ (*P4mm*) (c) NiO (*Fm-3m*)**

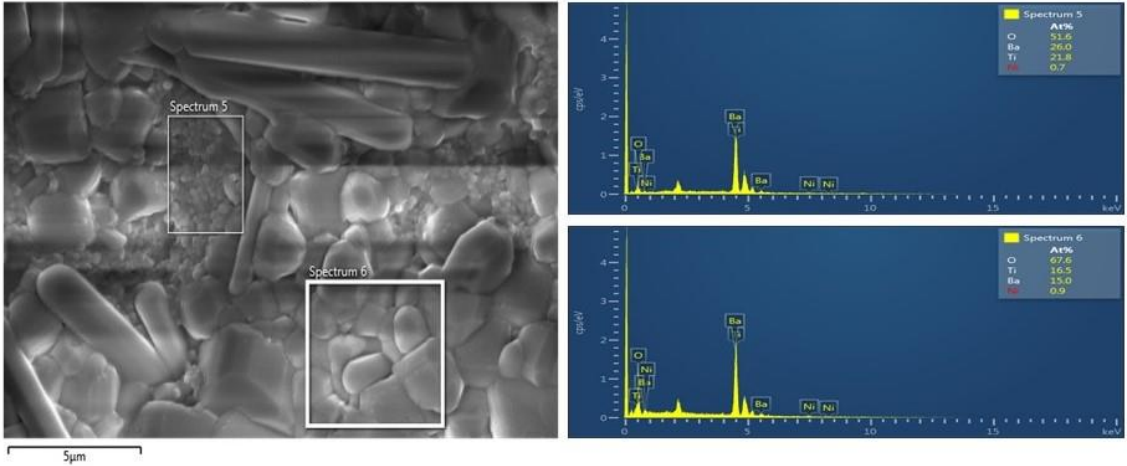

**Figure 4: Elemental analysis with EDX**

The Universal Dielectric Response model (UDR) is used to analyze the behavior of the dielectric material. According to the UDR model, the real part of the dielectric constant($\varepsilon'$) can be written as $\varepsilon' = \tan(s\pi/2)\, \sigma_0 f^{s-1}/\varepsilon_0$ where 's' and '$\sigma_0$' are frequency and temperature-dependent constants. In another form, $\varepsilon'*f = A(T)*f^s$ where $A(T) = \tan(s\pi/2)*\sigma_0/\varepsilon_0$ represents a temperature-dependent parameter. Again, which is, $\ln(\varepsilon'*f) = s*\ln(f) + \ln(A(T))$, so the plot between $\ln(f)$ vs. $\ln(\varepsilon'*f)$ will have the slope equal to 's'. In [Fig. 16] all the samples show a linear nature throughout the experimental frequency region. The linear fitting gives s values for S0, S1, S2, and S3 samples, respectively. The value of 's' decreases from 0.993 (S0) to 0.979 (S1), and 0.953 (S2), then shows a rising trend for S3 (0.976) shows a rising trend.

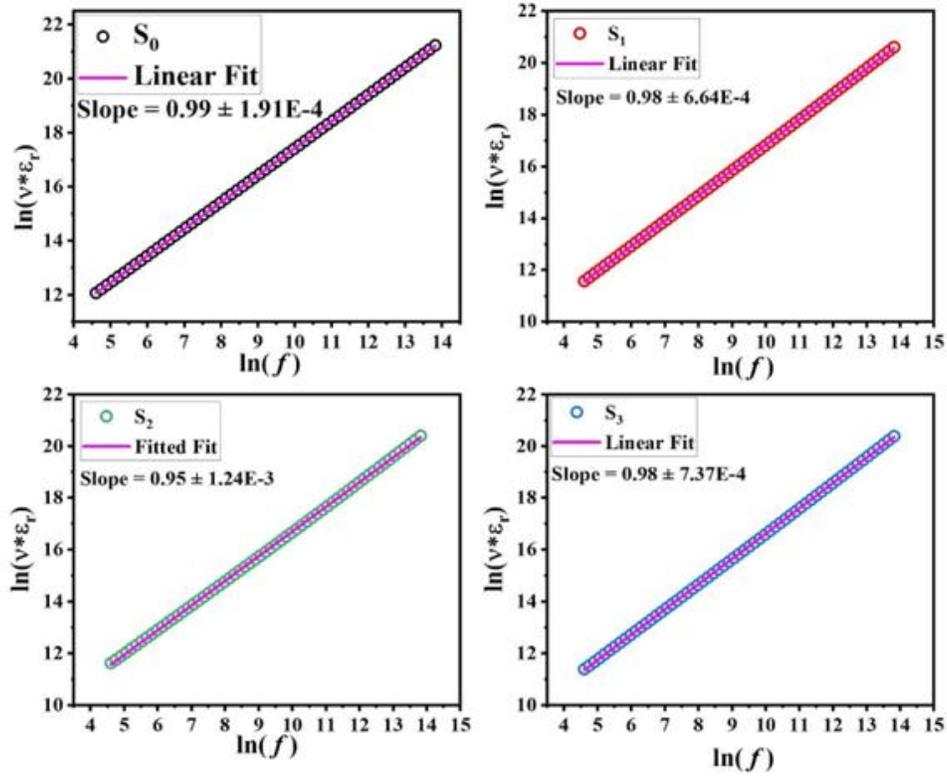

**Figure 5: Fitting plot for Universal Dielectric Response (UDR) Model**